# Peer Ratings in Massive Online Social Networks

**Dmitry Zinoviev**
*Mathematics and Computer Science Department, Suffolk University*
*dzinoviev@suffolk.edu*

**Abstract**

Instant quality feedback in the form of online peer ratings is a prominent feature of modern massive online social networks (MOSNs). It allows network members to indicate their appreciation of a post, comment, photograph, etc. Some MOSNs support both positive and negative (signed) ratings. In this study, we rated 11 thousand MOSN member profiles and collected user responses to the ratings. MOSN users are very sensitive to peer ratings: 33% of the subjects visited the researcher's profile in response to rating, 21% also rated the researcher's profile picture, and 5% left a text comment. The grades left by the subjects are highly polarized: out of the six available grades, the most negative and the most positive are also the most popular. The grades fall into three almost equally sized categories: reciprocal, generous, and stingy. We proposed quantitative measures for generosity, reciprocity, and benevolence, and analyzed them with respect to the subjects' demographics.

**Authors**

*Dmitry Zinoviev* is a professor of computer science at Suffolk University in Boston, MA. He conducts research on complex systems, including social networks and large computer and communications networks. His research interests in social networks are information diffusion, online vs offline friendship, and co-existence of semantic and social networks.

**Notes**

This research has been partially presented at SunBelt-2011, St Petersburg Beach, FL. The author is grateful to Prof. Lance Swanson (Suffolk University) and Dr. Alexander Semeonov (Higher School of Economics, Russia) for their valuable suggestions and to Mrs. Vy Duong for her great questions and small edits.

## 1. Introduction

Instant quality feedback in the form of online peer ratings is a prominent feature of modern massive online social networks (MOSNs). It allows users to indicate their appreciation of a post, comment, photograph, etc. The feedback can be provided either in the form of a free-text comment in a natural language (say, in English or in Russian) or in a discrete form by selecting an appropriate grading score from the list provided by the MOSN Web site (Crumlish, 2012).

The choice and the scope of the available grades varies widely between different MOSN sites. The simplest, single-valued mechanisms are used by Facebook (the "Like") button) and Google+ (the "+1" button). Single-valued ratings allow to express a positive attitude towards the rated resource—but not a negative one. Implicitly, resources with low cumulative rating may be thought of and often perceived as having a negative rating—but only compared to other higher rated resources.

Negative feedback can be introduced by implementing binary rating mechanisms known as "Thumbs Up/Thumbs Down." Such interfaces (found, say, on Pandora or Yahoo Answers) are often used to move the rated resource up or down a visual rating hierarchy. The binary ratings are more flexible than single-valued, but do not allow to express the degree of satisfaction or dissatisfaction.

Multivalued rating interfaces provide three or more stops. They can be balanced (ranging from "strong dislike" through "no opinion" to "strong like") or biased (ranging from "no opinion" to "strong like"). The stops are often represented as "stars." Just like with single-valued ratings, it is often implied that the "no opinion" stop of a biased interface has a negative connotation. Multivalued biased ratings can be found on Yahoo, EBay, and Amazon. We are not aware of major English-language social media sites using balanced multivalued ratings. These ratings are of particular interest to us.

It is not unusual for a posted rating to initiate a response reaction from the owner of the rated resource (post, photograph or comment). The case of social proximity between the grader and the graded, such as friendship or any other structural connection, implies repeated interaction between the parties and is relatively well studied (Kelman, 1974; Kollock, 1993; Dholakia, 2004; Leider, 2009). The reciprocal response is usually based on pro-social behavior and is altruistic—in the anticipation of future favors from the grader.

However, the situation is quite different in the case of a one-time encounter, when the rating has been posted not by a friend, relative, colleague or acquaintance, but by a stranger. Some of the few guidelines for posting the reciprocal feedback in this case are perceived enjoyment (a hedonistic motive to "be good," Li, 2011) and fear of noisy communication (when the owner of the graded resource suspects that the grader indeed is not a total stranger but someone familiar, whom the owner failed to recognize, Klapwijk, 2009).

In this paper, we experimentally study the one-time encounter situation and answer the following research questions:

- How many stops should the rating gauge have?
- How do MOSN members perceive the posted ratings?
- How do MOSN members respond to the posted ratings?

The rest of the paper is organized as follows: in Section 2, we explain the data acquisition process; in Section 3, we analyze and discuss the data; in Section 4, we conclude, and in Section 5, we present our plans for future work.

## 2. Method

Our study of peer ratings in massive online social networks (MOSN) is based on the data collected from "*Odnoklassniki*" ("Classmates"), the largest Russian-language social network with 148 mln users and 33 mln daily visits (Odnoklassniki 2011, 2013). "Odnoklassniki" is a "standard" MOSN Web site where members can post photographs and status updates, declare their basic demographics (such as age, gender, place of birth, education, and army service history), establish two-way friendship relationships, converse with other members, leave comments to photographs and status updates, etc. Most of these features are by default enabled for both friends and random visitors.

A distinctive feature of "Odnoklassniki" is its multivalued rating/grading system. Any image posted to the MOSN site, such as a profile picture, a personal album picture or a picture submitted to a thematic community (group), can be rated by authorized users on the scale from "1" (lowest, strong dislike) to "5" and "5+" (highest, strong like)[1]. The grade of "5+" requires a small monetary payment from the grader. The individual grades are visible only to the owner of the rated resources (photographs) and the grader herself.

The posted grades and comments can be easily removed by the resource owner. A MOSN member can also ban certain other members from posting grades or comments. We found both features essential for our experiment.

"Odnoklassniki" conveniently records the history of all profile visits; that is, any MOSN member can see who and when viewed her profile in the past[2].

We prepared a set of three MOSN site avatars (user profiles with pictures, Figure 1):

- an attractive female, 30 years old (F1),
- a bearded male, 40 years old (M1), and
- a beardless male, 30 years old (M2).

For each of these avatars Q, we experimented with 3,600 randomly selected MOSN members. For each member $U_i$, we recorded her age $A_i$ (the subject had to be at least 18 years old) and gender $G_i$. Then we posted a random grade $R_i^{in}$ (selected uniformly at random from all available grades between "1" and "5+") to the member's profile picture, recorded the member's response, and, in the case of a less-than-perfect grade of "1" through "4", apologized to her by leaving an appropriate comment. The data collection was administered manually by a human operator, produced 600 observations per grade per avatar, and took several weeks, so that we could collect both immediate and late responses. (Most of the time, the feedback either happened in the matter of minutes or did not happen at all.)

---

[1]Incidentally, the same grading scale is used in Russian public schools.

[2]A user can buy special service "*Nevidimka*"—"Invisible Man"—that makes her an invisible visitor, a visitor that does not leave traces. At the time when the data were collected, this service was not very popular.

The user's response, if any, consisted of one or more action from the list:

- visiting the avatar's profile ("visitor" behavior),
- posting a comment on the avatar's page ("commenter" behavior; implies visiting), and
- posting a reciprocal grade to the avatar's profile picture ("grader" behavior; implies visiting).

Our data set consists of 10,800 observations. There are 54% reportedly females and 46% reportedly males among the subjects. More statistics is shown in Table 1: visiting is the most common users' response, followed by reciprocal grading, followed by commenting. Male subjects are more active visitors than female subjects, while grading and commenting frequencies are more gender-agnostic.

We observed a remarkable stability of response rates across different age groups in the range of 18 to 75 years old. The mean age of the sampled population is 34 years.

## 3. Findings

Our experiment can be described in terms of the stimulus-response model (Wiener, 1961) where the subject $U_i$ is treated as a black box with certain properties that transforms a stimulus into a response:

$$[V, R^{out}, C] = f(R^{in}, Q, A, G) \quad (1)$$

In this equation:

- $R^{in} \in \{1.....5, 5+\}$ (the grade posted to the subject's profile picture) and $Q_i \in \{F1, M1, M2\}$ (the choice of the avatar) are the stimuli;
- $A \in [18, 75]$ (the age of the subject) and $G \in \{M, F\}$ (the gender of the subject) are the properties;
- $V \in \{True, False\}$ (whether the subject visited the avatar's profile), $R^{out} \in \{1.....5, 5+, \varnothing\}$ (the reciprocal grade posted by the subject to the avatar's profile picture, if any), and $C \in \{\varnothing, text\}$ (the comment left by the subject, if any) are the responses.

In the experiment, we control the stimuli and select the subjects to provide a reasonably uniform coverage of all combinations of stimuli and the subjects' properties.

### 3.1 Response Grades

Approximately 9% of all subjects posted a reciprocal grade to the avatars' profile pictures. The overall distribution of the reciprocally posted grades (the response grades) $R^{out}$ is shown in Figure 2. The distribution is sharply bimodal: the most popular ratings are "5" ("strong like") and "1" ("strong dislike"). Since the rating of "5+" has an associated monetary price and is not available to non-paying MOSN members, we treat it as indistinguishable from the vanilla "5" for the rest of the paper. The frequencies of the intermediate stops ("2", "3", and "4") and the differences between them are insignificant.

The distributions of the response grades, grouped by the stimulus grades, are shown in Figure 3. In each group, the paid rating of "5+" is dominated by its fee-free equivalent of "5," which supports our decision to lump the two ratings together.

Just like the overall distribution, the partial distributions for the lower ratings ("1", "2", "3", and somewhat even "4") are strongly bimodal, with the maximums at the lowest and highest fee-free ratings. The presence of the strong "5" component in the histograms 3a, 3b, and 3c suggests that the MOSN members are on average generous and often reward our avatars with disproportionately high reciprocal grades, even when being negatively rated. We will revisit this phenomenon later. It remains to be seen if the same generous behavior would be observed in the case of a rating scale with fewer stops.

For the higher stimulus ratings, the "dislike" maximum disappears at the expense of the "like" maximum, which becomes overwhelming.

Since the MOSN site users consistently ignore the intermediate grades, both on the lower and on the higher side, we believe that these grades are redundant and do not contribute to the efficiency of the peer rating mechanism. Future designers of social interfaces may limit the choices to the two signed extreme ratings, "strong dislike" and "strong like" (possibly with a neutral position of "no opinion").

To understand the subjects' reaction to the stimulus ratings better, we propose to split the numerical response into two components, reciprocity and generosity:

$$R^{out} = Rec[iprocity] \circ Gen[erosity] \quad (2)$$

Reciprocity is a social norm of in-kind responses to the behavior of others (also known as an "eye for an eye," Kollock, 1993)—in other words, a facility of mirroring the stimulus grade with the same grade.

Generosity is a habit of giving freely without expecting anything in return—that is, a facility of responding to a stimulus grade with a better grade. Perhaps somewhat counter-intuitively, we extend this concept to cover stinginess (lack of generosity). We define stinginess as negative generosity, a facility of responding to a stimulus grade with a worse grade.

In the next two Subsections, we explore reciprocity and generosity among the MOSN site users in detail.

### 3.2 Reciprocity

Let N be the total number of stimulus grades. We define reciprocity as the fraction of stimulus grades that were reciprocated with the identical response grade:

$$Rec = \frac{\Sigma_i \delta(R_i^{out} - R_i^{in})}{N} \quad (3)$$

Here, δ is the Kronecker delta function. The value of Rec is in the range from 0 (no posted grade is mirrored) to 1 (every posted grade is mirrored).

We calculated reciprocity values for each avatar and gender (Figure 4).

We used ANOVA to identify the factors that affect the corresponding measures. Modestly significant differences in mean reciprocity were found with respect to the interaction between the avatar's gender and the subject's gender ($p<0.1$):

- the female avatar F1 elicits more reciprocity from the subjects of the same gender and less reciprocity from the subjects of the opposite gender then the male avatars M1 and M2.

*3.3 Generosity*

The measure of reciprocity alone does not explain whether the subjects who are less likely to mirror the avatars' ratings, are less or more friendly towards the avatars.

We define generosity as the mean difference between the stimulus grade and the response grade:

$$Gen = \frac{\Sigma_i \left( R_i^{out} - R_i^{in} \right)}{N} \quad (4)$$

Unlike reciprocity, generosity is signed and ranges from -2.5 (when the subjects always give the lowest rating of "1" to the avatar's profile picture) to +2.5 (when the subjects always give the highest rating of "5").

We calculated generosity values for each avatar and gender (Figure 5).

Significant differences in mean generosity were found with respect to the avatar's age ($p<10^{-19}$), subject's age ($p<10^{-10}$), avatar's gender ($p<10^{-8}$), and interaction of the avatar's and subject's genders ($p<10^{-5}$). In particular:

- older subjects are less generous than younger subjects;
- the older avatar M1 elicits more generosity from the subjects;
- the male avatars M1 and M2 elicits more generosity from the subjects;
- male subjects are more generous to the male avatars and less generous to the female avatars than female subjects.

*3.4 Benevolence*

Reciprocity and generosity split the pair of stimulus and response grades into mirroring and gratification-antigratification components, making it possible to study them separately.

The final quantitative measure of the avatar-subject interaction is user comments. Each comment posted on the avatars' profile pages has been manually assigned to one of the three categories: negative (-1), positive (+1) or neutral (0). We define benevolence as the arithmetic mean value of a group of posted comments.

We calculated benevolence values for each avatar and gender (Figure 6).

Significant differences in mean benevolence were found with respect to the avatar's age ($p<10^{-5}$), subject's age ($p<10^{-5}$), avatar's gender ($p<10^{-5}$), and interaction of the avatar's and subject's genders ($p<10^{-5}$). Note that these are the same factors that controlled the generosity. However, their effect on benevolence is different:

- older subjects post more positive comments than younger subjects;
- the older avatar M1 elicits more negative comments from the subjects;
- the female avatar F1 elicits more positive comments from the subjects;
- male subjects post more positive comments to the female avatar F1 and more negative comments to the male avatars than female subjects.

In other words, benevolence is almost universally anticorrelated with generosity: better comments imply lower grades, and worse comments imply higher grades.

*3.5 Benevolence-Generosity Dilemma*

We call the phenomenon mentioned above the benevolence-generosity dilemma. To understand this effect, we looked at the group of 1,300 subjects who, in response to the stimulus, both left a comment and posted a reciprocal grade. For them we can calculate both generosity and benevolence. The ANOVA analysis shows that significant differences in mean benevolence exist with respect to generosity ($p<.0001$; the other significant factor is the avatar's age that has been already discussed above). The negative correlation between the two measures is small but stable (see also Figures 5 and 6). In fact, when the age factor is taken into account, the following observations can be made (overall across the avatars):

- younger subjects (18–30) are generous grade-wise, but post more negative comments;
- middle-aged subjects (30–40) are stingy and negative: they post lower grades and more negative comments;
- older subjects (40–75) post lower grades but more positive comments.

The generational difference may shed light on how MOSN site users perceive different peer rating and feedback mechanisms. At the moment, we do not have sufficient data to make an intelligent conclusion on this topic.

**4. Conclusion**

In this paper, we explored the mechanisms of multivalued peer ratings and feedback in massive online social networks (MOSNs). We initiated 10,800 interactions between researcher controlled avatars and actual users of "Odnoklassniki," the largest Russian-language MOSN. We collected and analyzed the outcomes of the interactions: history of subjects' visits to the avatars' profile pages, and comments and reciprocal grades posted by the subjects.

We observed that out of six possible ratings, only two (the lowest one, "strong dislike," and the highest fee-free one, "strong like") are used consistently and overwhelmingly, suggesting that a useful social interface does not need more than two stops on the rating scale.

We proposed the numerical measures of user feedback, both in terms of reciprocal grades and free-text comments: reciprocity, generosity, and benevolence—and analyzed their behavior with respect to the choice of the avatar and subject's age and gender. We noticed that of these factors, only the subject's gender does not have a strong influence on the proposed measures.

Finally, we discovered the benevolence-generosity dilemma—a "conservation law" that roughly preserves the overall emotional load of verbal (comments) and quantitative (grades) feedback. We observed that the younger subjects post more positive grades but more negative comments, while the older subjects do just the opposite. At the moment, we have no explanation of this behavior.

## 5. Future Work

We see two possible extensions of the project. First, the peer rating mechanism study could not be considered complete without a baseline calibration using a generic avatar—that is, an avatar without a profile picture or a non-human picture and with no posted demographics information. Another avatar notably missing from our set is an older female (as a counterpart to the older male). Collecting the feedback data for these two avatars is on the top of our list. Second, we consider applying game theoretical methods to study peer ratings as a case of a non-repeated (one-shot) game (Grüne-Yanoff, 2008). We expect that analyzing one-shot feedback from the perspective of MOSN members' utilities would help us understand their choice of the reciprocal grade.

## References

Crumlish, C. and Malone, E. (2009). *Designing Social Interfaces: Principles, Patterns, and Practices for Improving the User Experience* (Yahoo Press).

Dholakia, U.M., Bagozzi, R.P. and Pearo, L.K. (2004). A social influence model of consumer participation in network- and small-group-based virtual communities. *International Journal of Research in Marketing, 21*(3): 241–263.

Kelman, H.C. (1974). Further thoughts on the processes of compliance, identification, and internalization. *Social power and political influence*: 125–171.

Kollock, P. (1993). "An eye for an eye leaves everyone blind": Cooperation and accounting systems. *American Sociological Review, 58*: 768–786.

Leider, S., Möbius, M., Rosenblat, T. and Do, Q.A. (2009). Directed altruism and enforced reciprocity in social networks. *The Quarterly J. of Economics, 124*(4): 1815–1851.

Li, D. (2011). Online social network acceptance: a social perspective. *Internet Research, 21*(5): 562–580.

Klapwijk, A. and Van Lange, P. (2009). Promoting cooperation and trust in "noisy" situations: The power of generosity. *J. of Personality and Social Psychology, 96*(1): 83–103.

Odnoklassniki (2011). Retrieved in 2011 from http://www.odnoklassniki.ru.

Wikipedia, Odnoklassniki (2013). Retrieved 2013-06-23 from http://en.wikipedia.org/wiki/Odnoklassniki

Wiener, N. (1961). *Cybernetics: Or the Control and Communication in the Animal and the Machine* (MIT Press).

Wikipedia, Generosity (2013). Retrieved 2013-06-23 from http://en.wikipedia.org/wiki/Generosity.

Grüne-Yanoff, T. (2008). Game theory, Retrieved 2013-06-27 from http://www.iep.utm.edu/game-th/.

|   | Visitors | Graders | Commenters | Total |
|---|---|---|---|---|
| **M** | 18.4% | 10.4% | 2.6% | 31.4% |
| **F** | 14.7% | 10.9% | 2.3% | 27.9% |
| **Total** | 33.1% | 21.3% | 4.9% | 59.3% |

**Table 1.** Response rates by gender and category.

**CAPTIONS**

*Figure 1.* Avatars F1, M1, and M2.
*Figure 2.* Overall distribution of reciprocally posted grades.
*Figure 3.* Distributions of reciprocally posted grades for different stimulus grades $R^{in}$.
*Figure 4.* Reciprocity vs subject's age (best fit estimations); thick lines correspond to the male subjects, thin line—to the female subjects; solid lines correspond to the male avatar M1, dashed lines—to the female avatar F1, and dotted lines—to the male avatar M2.
*Figure 5.* Generosity vs subject's age. See Figure 4 for the legend.
*Figure 6.* Benevolence vs subject's age. See Figure 4 for the legend.

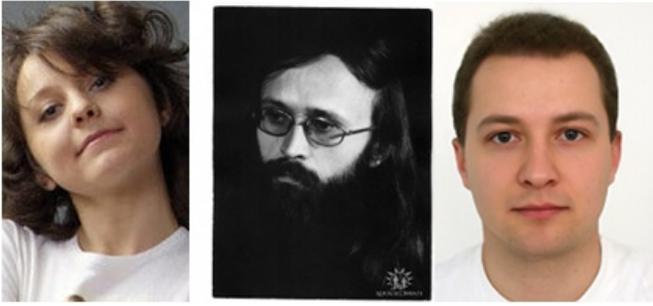

*Figure 1.* Avatars F1, M1, and M2.

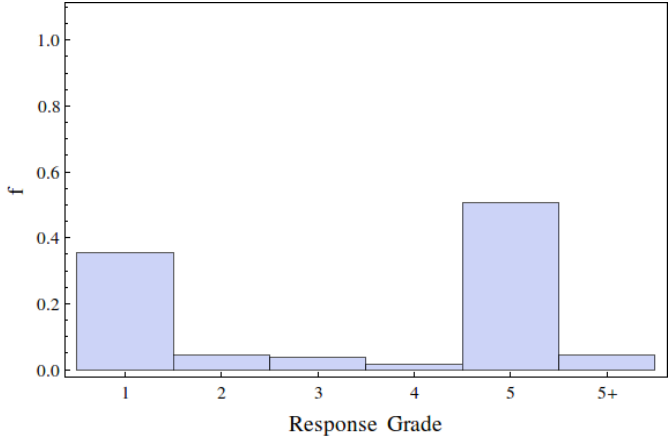

*Figure 2.* Overall distribution of reciprocally posted grades.

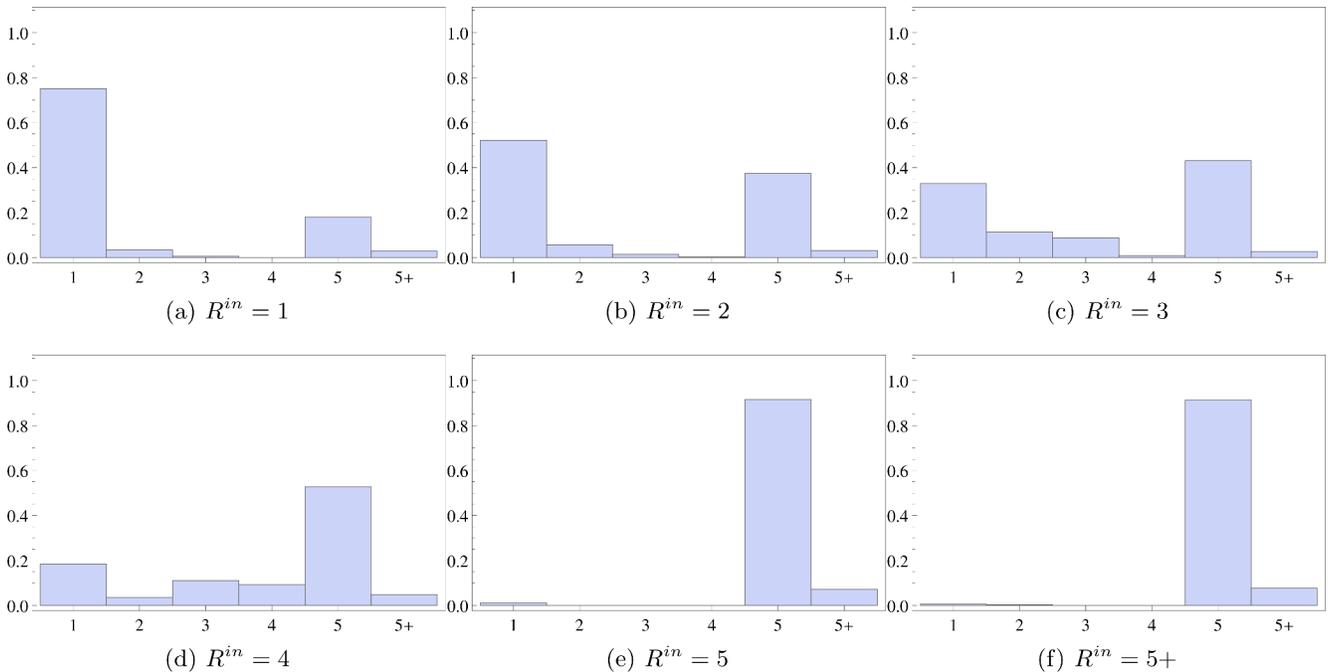

*Figure 3.* Distributions of reciprocally posted grades for different stimulus grades $R^{in}$.

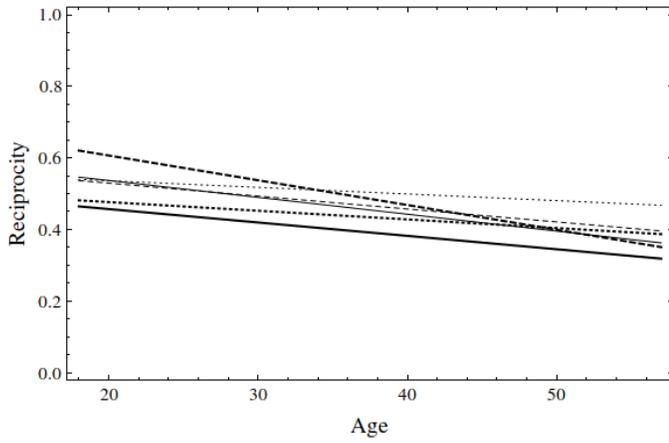

*Figure 4.* Reciprocity vs subject's age (best fit estimations); thick lines correspond to the male subjects, thin line—to the female subjects; solid lines correspond to the male avatar M1, dashed lines—to the female avatar F1, and dotted lines—to the male avatar M2.

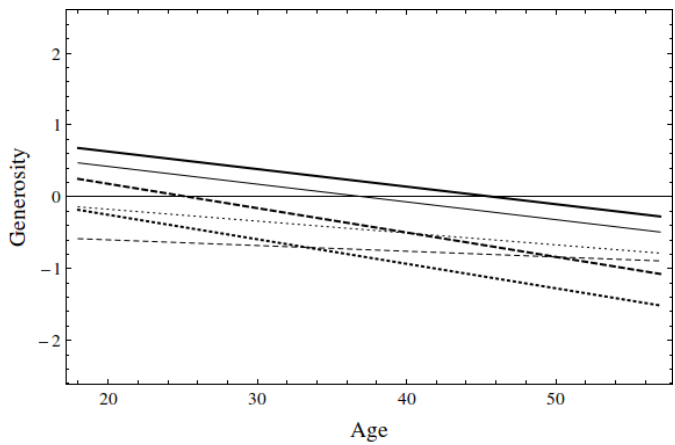

*Figure 5.* Generosity vs subject's age. See Figure 4 for the legend.

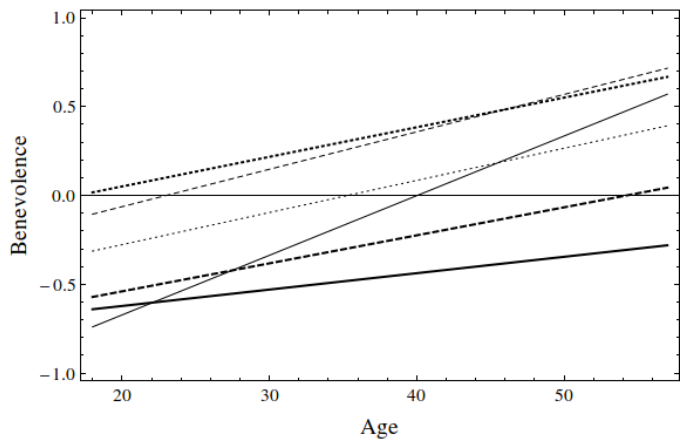

*Figure 6.* Benevolence vs subject's age. See Figure 4 for the legend.